\documentclass[slac_one]{revtex4}
\usepackage{graphicx}
\usepackage{fancyhdr}
\def\ie{{\it i.e.}}
\def\eg{{\it e.g.}}
\def\etc{{\it etc}}
\def\etal{{\it et al.}}

\def\to{\rightarrow}

\def\mpl{\ifmmode \overline M_{Pl}\else $\overline M_{Pl}$\fi}

\begin{document}

\rightline{\vbox{\halign{&#\hfil\cr
%&DRAFT\cr
&SLAC-PUB-10753\cr
&September 2004\cr}}}

\title{Pedagogical Introduction to Extra Dimensions}


\author{Thomas G. Rizzo}
\affiliation{SLAC, Stanford, CA 94025, USA}

\begin{abstract}
Extra dimensions provide a new window on a number of problems faced by the 
Standard Model. The following provides an introduction to this very 
broad subject aimed at experimental graduate students and post-docs based on a 
lecture given at the 2004 SLAC Summer Institute.
\end{abstract}

\maketitle

\thispagestyle{fancy}

\section{Introduction: Why Study Extra Dimensions?}

Most physicists agree that some form of New Physics (NP) must exist beyond the 
Standard Model(SM)--we simply don't know what it is yet. Though there are many 
prejudices based on preconceived ideas about the form this NP may take, it 
will be up to experiments at future colliders, the LHC and ILC, to reveal 
its true nature. While we are all familiar with the list of these theoretical 
possibilities one must keep in mind that nature may prove to be more 
creative than we are and that something completely unexpected may be 
discovered. After all, we certainly have not yet explored more than a fraction 
of the theory landscape. 

One now well-known possibility is that extra spatial dimensions will begin 
to show themselves at or near the TeV scale. Only a few years ago not many 
of us would have thought this even a remote possibility, yet the discovery 
of extra dimensions(EDs) would produce a fundamental change in how we view 
the universe. The study of the physics of TeV-scale EDs that has taken place 
over the past few years has its origins in the ground breaking work of 
Arkani-Hamed, Dimopoulos and Dvali(ADD){\cite {ADD}}. Since that time EDs 
has evolved from a single idea to a new paradigm with many authors employing 
EDs as a tool to address the large number of outstanding issues that remain 
unanswerable in the SM context. This in turn has lead to other 
phenomenological implications which should be testable at colliders and 
elsewhere. 
A (very) incomplete list of some of these ideas includes, \eg,   
\begin{itemize} 

\item addressing the hierarchy problem{\cite {ADD,RS}}

\item producing electroweak symmetry breaking without a 
Higgs boson{\cite {nohiggs}}

\item the generation of the ordinary fermion and neutrino mass hierarchy, the 
CKM matrix 
and new sources of CP violation{\cite {flavor}}

\item TeV scale grand unification or unification without SUSY while 
suppressing proton decay{\cite {unif}}

\item new Dark Matter candidates and a new cosmological perspective{\cite {ued,bine}} 

\item black hole production at future colliders as a window on quantum 
gravity{\cite {bh}}

\end{itemize}
This list hardly does justice to the wide range of issues that have been 
considered  in the ED context. Clearly a discussion of all these ideas is 
beyond the scope of the present work and only some of them will be briefly 
considered in the text which follows. However it is clear from this list 
that ED ideas have found their way into essentially every area of interest 
in high energy physics which certainly makes them worthy of detailed study. 

Of course for many the real reason to study extra dimensions is that they 
are fun to think about and almost always lead to surprising and unanticipated 
results.

\section{Thinking About Extra Dimensions}

Most analyses of EDs are within the context of quantum field theory. Can we 
learn anything about EDs from `classical' considerations and some general 
principles without going into the complexities of field theory? Consider 
a massless 
particle moving in 5D `Cartesian' co-ordinates and assume that 5D Lorentz
invariance holds. Then the square of the 5D momentum for this particle is 
given by $p^2=0=g_{AB}p^Ap^B=p_0^2-${\bf p}$^2\pm p_5^2$ where $g_{AB}=
diag(1,-1,-1,-1,\pm 1)$ is the 5D metric tensor (\ie, defined by the 
invariant interval $ds^2=g_{AB}dx^Adx^B$), 
$p_0$ is the usual particle 
energy, {\bf p}$^2$ is the square of the particle 3-momentum and $p_5$ is 
its momentum along the 5th dimension. (Note that the indices $A,B$ run over 
all 5D. We will sometimes denote the 5th 
dimension as $x_5$ and sometimes just as $y$.) The `zero' in the equality  
above arises from the fact that the particle is assumed to be 
massless in 5D. Note that a priori we do not know 
the sign of the metric tensor for the 5th dimension but as we will now see 
some basic physics dictates a preference; the choice of the +(-) sign 
corresponds to either a time- or space-like ED. We can re-write the equation 
above in a more traditional form 
as $p_0^2-${\bf p}$^2=p_\mu p^\mu =\mp p_5^2$ and we recall, for 
all the familiar particles we know of which satisfy 4D Lorentz invariance,
that $p_\mu p^\mu=m^2$, which is just the square of the 
particle mass. (Note that Greek indices will be assumed to run only over 4D here.) 
Notice if we choose a time-like extra dimension that the 
sign of the square of the mass of the particle will appear to be 
{\it negative}, \ie, the particle is a {\it tachyon}. Tachyons are well known to 
be very dangerous in most models, even classically, as they can cause severe 
causality problems{\cite {tach}}--something we'd like to avoid in 
any theory--provided they interact with SM particles. This implies that we 
should pick the space-like solution. Generally, it turns out that to 
avoid tachyons appearing in our ED theory we must always choose EDs to be 
space-like and therefore we assume there will always be only one time dimension 
even though we could all use some extra time.{\footnote {See however 
{\cite {dvali}} for a discussion of time-like EDs.}}.  

Now lets think about a real massless scalar field in a flat 5D-space  
(assuming a space-like ED!) which is a solution of the 5D Klein-Gordon equation: 
($\partial_A \partial^A)\Phi=(\partial_\mu \partial^\mu-\partial_y^2)
\Phi(x,y)=0$, where $y$ here represents the extra dimension. 
We can do a fast and dirty trick by performing something 
like separation of variables, \eg, take $\Phi= \sum_n \chi_n(y) \phi_n(x)$ 
and plug it into the Klein-Gordon equation
above giving us $\sum_n(\chi_n \partial_\mu \partial^\mu 
\phi_n-\phi_n \partial_y^2 \chi_n)=0$. Now we note that if $\partial_y^2 \chi_n
=-m_n^2\chi_n$, we obtain a set of equations that appears like 
$\sum_n \chi_n (\partial_\mu \partial^\mu+m_n^2)\phi_n=0$ which looks like 
an infinite set equations for a collection 
of distinct 4D scalar fields $\phi_n$ with masses 
$m_n$. This bunch of states with different masses is called a Kaluza-Klein(KK) 
tower. Note that we labeled the states by a set of integers ($n$) so that 
the levels are discrete; we could just as well have replaced the sum by an 
integral 
and treat $n$ as a continuous variable. The difference between these two 
possibilities and the link to the nature of the 5D space and the associated 
boundary conditions will be made clear below.

Now we have pulled a bit of a fast one in performing this quick and dirty 
analysis so let us 
return and do a somewhat better job; we will still assume, however, that 
$n$ is a distinct integer label for reasons to be clarified below. 
Let us start from the action (\ie, the integral of the Lagrangian) for 
the massless 
5D scalar assuming $y_1\leq y \leq y_2$ with $y_{1,2}$ for now arbitrary:

\begin{equation}
S=\int d^4x \int_{y_1}^{y_2} ~dy ~{1\over {2}}\partial_A \Phi \partial^A \Phi  \,.
\end{equation}
Now we recall $\partial_A \Phi \partial^A \Phi= \partial_\mu \Phi \partial^\mu 
\Phi -\partial_y \Phi \partial_y \Phi$ and substitute the decomposition $\Phi= 
\sum_n \chi_n(y) \phi_n(x)$ as above. Then the integrand of the action 
becomes a double sum proportional to $\sum_{nm}[ \chi_n\chi_m \partial_\mu 
\phi_n \partial^\mu \phi_m -\phi_n\phi_m \partial_y \chi_n \partial_y \chi_m]$. 
This appears to be a mess but we can `diagonalize' this equation in a few steps. 
First if we choose to orthonormalize the $\chi_n$ such that 

\begin{equation}
\int_{y_1}^{y_2} ~dy ~\chi_n \chi_m=\delta_{nm}  \,,
\end{equation}
then the kinetic term of the $\phi_n$ (the first one in the bracket above) reduces 
to a single sum after $y$ integration becoming simply $\sum_n \partial_\mu \phi_n 
\partial^\mu \phi_n$. This is essentially just sum of the kinetic terms for a set of 
distinct 4D scalars. To handle the second term in the brackets we integrate 
by-parts and note that {\it if} we take the boundary conditions to be 
\begin{equation}
\chi_m \partial_y \chi_n |_{y_1}^{y_2}=0\,,
\end{equation}
and also require that 
\begin{equation}
\partial_y^2 \chi_n=-m_n^2\chi_n \,,
\end{equation}
as above, we can integrate the entire action over $y$ and obtain an effective 
4D theory  
\begin{equation}
S=\int d^4x ~{1\over {2}} \sum_n[\partial_\mu \phi_n \partial^\mu \phi_n-
m_n^2 \phi_n^2]  \,,
\end{equation}
which is just the sum of the actions of the independent 4D scalars labeled by 
$n$ with masses $m_n$, \ie, the KK tower states. One sees that in this 
derivation it was important for 
the above boundary conditions(BCs) to hold in order to obtain this result.
It is important to note that in fact the various masses 
we observe in 4D correspond to (apparently) quantized values of $p_y$ for the 
different $\phi_n$.

The fields $\chi_n$ can be thought of as the wave functions of the various 
KK states in the 5th dimension and in this simple, flat 5D scenario are 
simple harmonic functions: $\chi_n=A_n e^{im_ny}+B_n e^{-im_ny}$. What are the 
$m_n$? What are $A_n$ and $B_n$? To say more we must discuss BCs a bit further. 
In thinking about BCs in this kind of model it is good to recall one's 
experience with one-dimensional Quantum Mechanics(QM) 
that we all learned (too) many 
years ago. First recall the Schr\"odinger Equation for a free particle 
moving along the $x$ direction. It has the same form as Eq.4 above and since 
the $x$ direction is infinite, \ie, {\it noncompact}, the solution is just 
$\psi \sim A'e^{ipx}+B'e^{-ipx}$ where $p$ is the particle 
momentum which can take on an 
infinite set of continuous values. We say that in this case the momentum $p$ 
is not quantized and this is due to the fact that the space is noncompact. 
Now let us consider a slightly different problem, a particle in 
a box, \ie, a situation where the potential is zero for $0 \leq y \leq \pi L$ 
but is infinite elsewhere so that the wavefunction vanishes outside this 
region. Since the physical region is of a finite size it is called {\it compact}. 
We know that the solution inside the box 
takes the same general form as does the case of a free particle or $\chi_n$ 
above but it must also vanish at the boundaries. These BCs tell us $A'$ and $B'$ 
so that the 
solutions actually takes the form $\sim \sin ny/L$ and that the momenta are 
quantized, \ie, $p=n/L$ with $n=1,2,...$. 
Clearly these two situations are completely 
analogous to having a 5th dimension which is either infinite (\ie, noncompact) 
or finite (\ie, compact) in size. {\it Under almost all circumstances} we 
assume that extra dimensions are compact. For a flat 5th dimension of length 
$\pi L$ the analysis above tells us that the KK masses are given by 
$m_n=n/L$, \ie, the masses are  large if the size of the extra dimension is 
small. Perhaps it is natural to think that the reason that we have not 
seen EDs is that they are 
very small and the corresponding KK states are then too massive to be 
produced at colliders. In fact, the observation of KK excitations is the 
hallmark of EDs. It is interesting to observe that there are 
no solutions in the `particle in a box' example  
corresponding to massless particles, \ie, those with $n=0$, the so-called 
zero modes.

There are other sorts of BCs that can be important. In QM we also examined the 
case of  
a particle moving on a circle where angular momentum is quantized. In the 5D 
case we can imagine that the 5th dimension is curled up into a circle, 
$S^1$--a one dimensional sphere of radius $R$--so 
that the points $y=-\pi R$ and $y=\pi R$ are the same, \ie, we have periodic 
boundary conditions. Here the KK masses 
are given by $m_n=n/R$ so that we still see the correlation between the KK 
masses and the inverse size of the extra dimensions but the solutions take 
the form of $ A_n\cos ny/R +B_n\sin ny/R$ with $n=0,1,2,...$. Note that here 
a massless mode does exist due to the periodicity of the BCs. We can change this 
solution slightly by imagining defining a parity operation on the interval 
$-\pi R \leq y \leq \pi R$ which maps $y\to -y$. There are now 2 special points 
on this interval, call the fixed points, which are left invariant by this $Z_2$ 
operation when combined with the translation $y\to y+2\pi R$ and the periodicity 
property; these are the points $y=0,\pi R$. The eigenfunctions of our 
wave equation must now respect the discrete, $Z_2$, parity symmetry so that 
our solutions are {\it either} $Z_2$-even, $\sim \cos ny/R$ or $Z_2$-odd,  
$\sim \sin ny/R$. Note that only $Z_2$-even 
states have a zero-mode amongst them. This geometry 
is called $S^1/Z_2$ and is the simplest example of an {\it orbifold}, a 
manifold with a discrete symmetry that identifies different points in the 
manifold, here $y$ and $-y$.  The $S^1/Z_2$ orbifold is of particular 
interest in model building as we will see below. By the way, we note that 
all of the BCs of interest to us above are such as to satisfy the conditions 
following from Eq.3.

So far we have seen that a 5D scalar field decomposes into a tower of 4D 
scalars when going from the 5D to 4D framework. What about other 5D fields? 
In a way we have encountered a somewhat similar question when we first 
learned Special Relativity, \ie, how do 3-vectors and scalars get embedded 
into 4D fields? Just as a 4-vector contains a 3-vector and a 3-scalar, one 
finds that, \eg, 
a 5D massless gauge field (which has 3 polarization states!) 
contains two KK towers, one corresponding to a 4D gauge field (with 2  
polarization states) and the other to a 4D scalar field. In fact in 
(4+n)-dimensions a gauge field will decompose into a 4D gauge KK tower plus 
$n$ distinct scalar towers. At this point a subtlety exists. We know from 
our previous discussion that KK tower states above the zero mode are massive. 
How can there be massive 4D gauge fields with only 2 transverse 
polarization states? 
Consider for simplicity the 5D case. It turns out that we can make a gauge 
transformation to eliminate the scalar KK tower fields and have them `eaten' 
by the gauge field--in a manner similar to the Higgs-Goldstone 
mechanism--becoming their longitudinal components. 
In a way this is a geometric 
Goldstone mechanism. The scalar KK tower is then identified to be  
just the set of Goldstone bosons eaten by the gauge fields to acquire masses. 
Thus in the {\it unitary} or physical gauge the massless 5D gauge field 
becomes a massive tower of 4D gauge fields. In (4+n)-dimensions only one 
linear combination of the scalars is eaten so that a  massless 
(4+n)-dimensional gauge field 
produces a massive tower of 4D gauge fields together with 
$n-1$ scalar towers in the unitary gauge. 

To see how this 5D decomposition for gauge fields 
works in practice consider a massless 5D gauge field with 
the 5th dimension compactified on $S^1/Z_2$. In the first step in the KK 
decomposition one can show that the 2-component vector KK's are $Z_2$-even 
with 5D wavefunctions like $\sim \cos ny/R$, whereas the KK scalars are 
$Z_2$-odd with wavefunctions like $\sim \sin ny/R$. Note the absence of 
an $n=0$ scalar mode due to the $Z_2$ orbifold symmetry. Once we employ the 
KK version of the Higgs-Goldstone trick all the $n>0$ gauge KK tower fields 
become massive 3-component gauge fields with $m_n=n/R$, having eaten their 
scalar partners. However, the zero mode remains massless, \ie, 
there was no Goldstone boson for it to eat. In 4D the masslessness of the zero 
mode tells us that gauge invariance has not been broken. We see that 
orbifold BCs are useful at generating massless modes and maintaining gauge 
invariance. One can translate these orbifold BCs for the gauge 
fields into the simple relations $\partial_y A^\mu_n|=0, ~A^5_n|=0$ for all 
$n$ and where `$|$' stands for a boundary. Interestingly the physics changes 
completely if we change the BCs in this case. Instead of an orbifold, 
consider compactifying on a line segment or interval, $0\leq y\leq \pi R$, and 
taking as BCs $A^\mu_n=\partial_y A^5_n=0$ at $y=0$ and 
$\partial_y A^\mu_n=A^5_n=0$ 
at $y=\pi R$. Here one now finds that $A^\mu \sim \sin m_ny/R$ and no 
massless zero mode exists. After the Higgs-Goldstone trick {\it all} the 
4D gauge KK tower fields become massive (leaving no remaining scalars) with 
$m_n=(n+1/2)/R$ implying that gauge invariance is now broken. Here we see a 
simple example that demonstrates that we can use BCs to break gauge 
symmetries. The possibility that such techniques can be successfully employed 
to break the symmetries of the SM without the introduction of 
fundamental Higgs fields 
has been recently extensively discussed in the literature{\cite {nohiggs}}. 

How do other higher dimensional fields decompose? As the simplest example 
consider the gravitational field in 5D, represented as by symmetric tensor 
$h^{AB}$ which decomposes as 
$h^{AB}\rightarrow (h^{\mu\nu},h^{\mu 5},h^{55})_n$ when going to 4D; as 
before $n$ is a KK tower index. If we 
compactify on $S^1/Z_2$ to keep the zero mode $h^{\mu\nu}_0$ massless (and 
which we'll  identify as the ordinary graviton of General Relativity) then for 
$n>0$ all of the $h^{\mu 5}_n$ and $h^{55}_n$ fields get eaten to generate 
a massive KK graviton tower with fields that have 5 polarization states (as 
they should).  For $n=0$ there is no $h^{\mu 5}_0$ to eat due to the orbifold 
symmetry yet a massless $h^{55}$ scalar remains in addition to 
the massless graviton. 
This field is the {\it radion}, which corresponds to an fluctuation of the 
size of the extra dimension, and whose vacuum value must be stabilized by 
other new physics to keep radius the extra dimension stable{\cite {gw}}. The 
physical spectrum is then just the radion and the graviton KK tower in this 
case which, as we will see below, corresponds to what happens in the RS model. 

In terms of manifolds on which to compactify higher dimensional fields it is 
easy to imagine that as we go to higher and higher dimensions the types of 
manifolds and the complexity of the possible orbifold symmetries grows rapidly. 
Typical manifolds that are most commonly considered are torii, $T^n$, which 
are simply products of circles, and spheres, $S^n$. 

Let us now turn to a few important representative ED models.

\section{Large Extra Dimensions}

The Large Extra Dimensions scenario of Arkani-Hamed, Dimopoulos and 
Dvali(ADD){\cite {ADD}} was proposed as a potential solution to the hierarchy 
problem, \ie, the question of why the (reduced) Planck scale, 
$\mpl \simeq 2.4 \cdot 10^{18}$ GeV, is so much larger than the weak scale 
$\sim 1$ TeV. ADD propose that we (and all other SM particles!) live on an  
assumed to be rigid 4D hypersurface (sometimes called a wall or brane). On 
the other hand gravity is allowed to propagate in a 
(4+n)-dimensional `bulk' which is, \eg, an $n-$torus, $T^n$. This brane is 
conveniently located at the origin in the EDs, \ie, {\bf y}=0. 
Gauss' Law then tells us 
that the Planck scale we measure in 4D, $\mpl$, is related to the true 
(4+n)-dimensional fundamental 
scale, $M_*$, that appears in the higher dimensional action, via the relation
\begin{equation}
\mpl^2=V_nM_*^{n+2}\,,
\end{equation}
where $V_n$ is the volume of the $n$-dimensional compactified space. 
$M_*$ (sometimes denoted as $M_D$ in the literature) can be thought of as the 
{\it true} Planck scale since it appears in the higher dimensional action which 
is assumed to describe General Relativity in (4+n)-dimensions. Is is possible 
that $M_*$ could be $\sim$ a few TeV thus `eliminating' the hierarchy problem? 
Could we have been fooled in our extrapolation of the behavior of gravity from 
what we know up to the TeV scale and beyond and that gravity, becomes strong at 
$M_*$ and not at \mpl? To get an idea whether this scenario 
can work at all we need to 
get some idea about the size of $V_n$, the volume of the compactified space. 
As a simple example, and the one most often considered in the literature, 
imagine that this space is a torus $T^n$ all of whose radii are equal to $R$. 
Then it is easy to see that $V_n=(2\pi R)^n$; knowing \mpl ~and assuming 
$M_*\sim$ a few TeV we can estimate the value of $R$. Before we do this we need 
to think about how gravity behaves in EDs. 

\begin{figure}[htbp]
\centerline{
\includegraphics[width=12.0cm,angle=0]{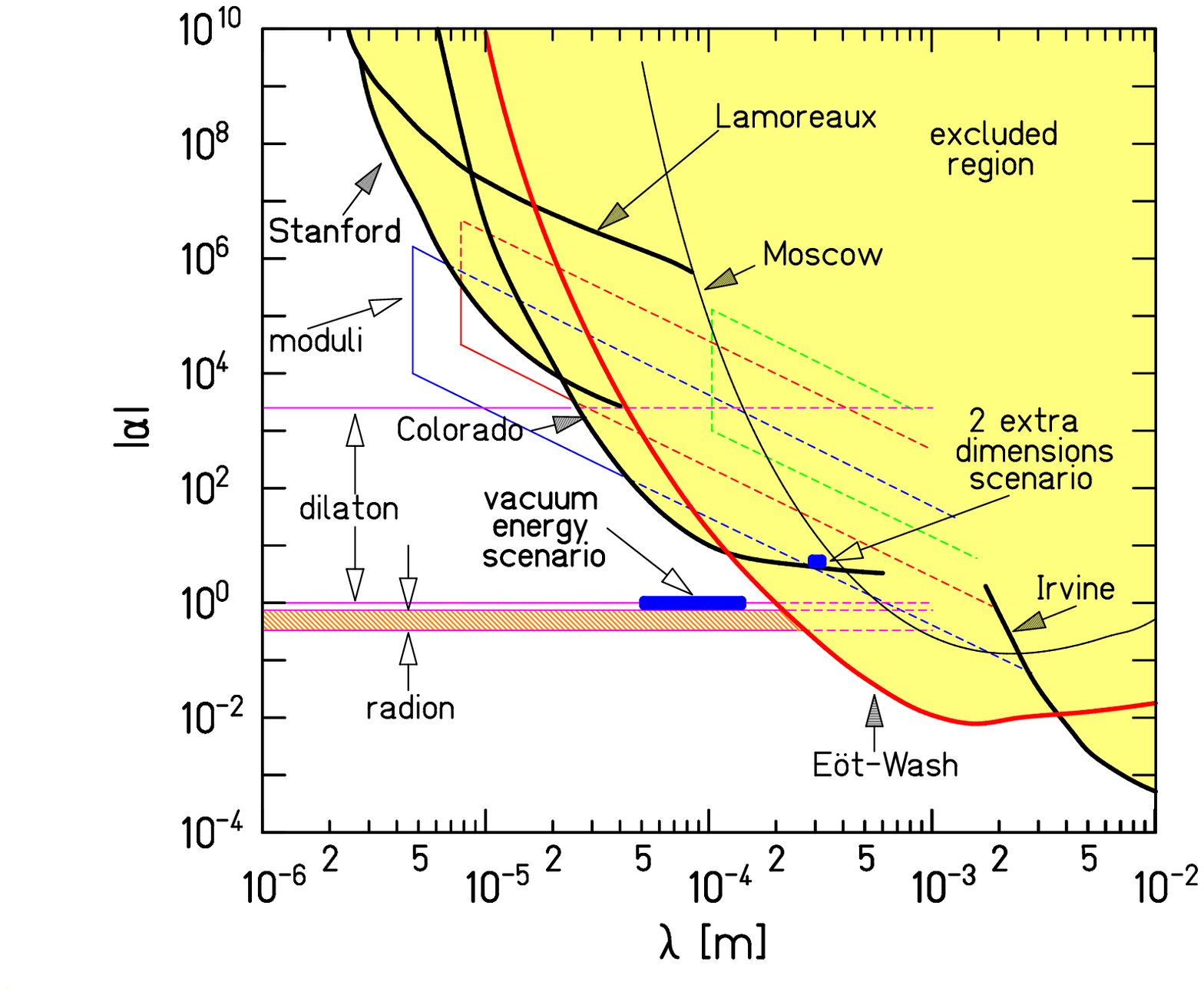}}
\vspace*{0.0cm}
\caption{Regions in the $\alpha-\lambda$ plane excluded by table top searches for 
deviations from Newtonian gravity from Adelberger \etal{\cite {adel}}. The ADD 
prediction with $n=2$ and $M_*=1$ TeV is also shown.}
\label{fig1}
\end{figure}

If one considers two masses separated by a distance $r$ in $(n+4)$-dimensions  
the force of attraction 
now depends on the relative magnitudes of $r$ and the compactification radius $R$. 
To see this first imagine $r>>R$; in this case the extra dimensions are 
essentially invisible and to all appearances the space looks to be 4D. Then we 
know that $F_{grav} \sim 1/r^2$ thanks to Newton. However, in the opposite 
limit $r<<R$ the effects of being in a full (4+n)-dimensional space will 
become obvious; we don't even realize that the ED is compactified. 
Either via Gauss' Law or by recalling the nature of the 
solution to the inhomogeneous Laplace's equation in EDs one finds that now 
$F_{grav} \sim 1/r^{2+n}$. Clearly one will start to see significant deviations 
from Newtonian gravity once $r\sim R$ so 
that $R$ cannot be very large. Let's 
assume $n=1$; then we can solve the  equation above and obtain $R\sim 10^{8}$ m. 
This is a scale of order the Earth-Sun distance over which we know Newton's Law 
holds very well; thus $n=1$ is excluded. Fortunately the size of $R$ 
decreases rapidly as $n$ increases; for $n=2$ one obtains $R\sim 100 \mu$m 
which is close to the limit of current table top 
experimental searches{\cite {adel}}  
for deviations from Newton's Law of Gravity. These are summarized 
in Fig.~{\ref {fig1}} from the work of Adelberger \etal{\cite {adel}}. 
Note that these deviations 
from Newtonian gravity are conventionally parameterized by adding a Yukawa-type 
interaction of relative 
strength $\alpha$ and scale length $\lambda$ to the usual Newtonian 
potential. In the figure the deviations 
expected in the $n=2$ scenario are shown assuming $M_*=1$ TeV; the bounds from 
the data tell us that $M_*>$ a few TeV in this case.

If $n$ is further increased $R$ 
becomes way too small to probe for direct deviations from the $1/r^2$ Law. It is 
interesting to note that for $n=2$, which is already being constrained by table top 
measurements, $R^{-1} \sim 10^{-4}$ eV telling us that we have not probed 
gravity directly beyond energy scales of this magnitude. This ignorance 
is rather amazing but it is what allows the large parameter space in which the ADD 
model successfully functions. From this discussion it appears that the ADD model 
will work as long as $n\geq 2$ with $n=2$ being somewhat close to the boundary 
of the excluded regime. As we will see below the naive $n=2$ case is highly 
disfavored by other measurements though larger values of $n$ are much more weakly 
constrained. How large can $n$ be? If we believe in superstring theory at high 
scales then we can expect that $n\leq 6$ or 7. A priori, however, there is no reason 
not to consider larger values in a bottom-up approach. 
It is curious to note that when  $n\simeq 30$ one has  
$M_*R \sim 1$ which is perhaps what we might expect based on naturalness assumptions;  
for any $n\sim 15$ or less, $R^{-1}<<M_*\sim$ a few TeV. This point is important for 
several reasons. ($i$) One may ask why we required that the SM fields remain on the 
brane. If the SM or any part thereof were in the bulk, those fields would have KK 
towers associated with them. Since the masses of these KK fields would be of order 
$\sim 1/R$ as discussed above {\it and} we have not experimentally observed any KKs  
of the SM particles at collider so far, we must have $1/R \geq 100$ GeV or so. For any 
$n\leq 10$ it is clear that this condition cannot be satisfied. So if we believe in 
strings the SM fields must remain on the wall. ($ii$) Since the gravitons {\it are} 
bulk fields their KK masses are given by $m_{KK}^2=\sum_{i=1}^n l_i^2/R^2$ where 
$l_i$ is a integer labeling the KK momenta for the $i^{th}$ ED. As we noted above,  
for not too large values of $n$ these masses are generally very small compared to the 
1 TeV or even 100 GeV scale. This will have important phenomenological implications 
below. 

If $n=6$ or 7 in string theory why don't we just assume this value? Consider a small 
variation on a theme. So far we have assumed that all of the ED compactification 
radii are equal; this need 
not be the case. Assume there are $n$ EDs but let $n-p$ of them have radius 
$R_1$ and $p$ of them radius $R_2$. Then from above we have  
\begin{equation}
\mpl^2= (2\pi)^n  R_1^{(n-p)} R_2^p M_*^{(n+2)}\,.
\end{equation}
Now imagine that $R_2^{-1}\sim M_*$; then we'd have instead  
\begin{equation}
\mpl^2 \sim R_1^{(n-p)} M_*^{(n-p)+2}\,,
\end{equation}
\ie, it would {\it appear} that we really only have $n-p$ large dimensions. The keyword 
here is `large', the $p$ dimensions are actually `small' of order $\sim$ TeV$^{-1}$ 
and not far from the fundamental scale in size. 
Thus there could be, \eg, 7 EDs as suggested by strings but only 4 of them are 
large. If any SM field lived {\it only} in these $p$ small EDs that would be (at least 
superficially)  
experimentally acceptable since their KK masses would be $\sim$ TeV and as of yet 
these KKs would be 
unobserved at colliders.  Since the SM fields can live in these TeV size EDs we will 
see below this the possibility is quite popular{\cite {aton}} for model building purposes. 

To proceed further we need to know what these KK graviton states do, \ie, how they 
interact with the SM fields on the wall. A derivation of the Feynman rules for the 
ADD model is beyond the scope of the present talk but can be found in 
Ref.{\cite {hlzgrw}} with some elementary 
applications discussed in Ref.{\cite{jlhmp}}. 
A glance at the Feynman rules tells us several things: ($i$) all of the states in 
the graviton KK tower couple to SM matter on the wall with the same strength as the 
ordinary zero-mode graviton, \ie, to lowest order in the coupling  
\begin{equation}
{\cal L}=-{1\over {\mpl}} \sum_n G^{\mu\nu}_n T_{\mu\nu}\,,
\end{equation}
where $G^{\mu\nu}_n$ are the KK graviton fields in the unitary gauge and 
$T_{\mu\nu}$ is the stress-energy tensor of the SM wall fields. ($ii$) Since there 
are at least 2 EDs  
we might expect that the vector fields (or some remaining 
combination of them) $G^{\mu i}_n$, where $i=1,...,n$,   
would couple to the SM particles. It turns out that such couplings are absent by 
symmetry arguments since the SM wall resides at {\bf y}=0. ($iii$) Similarly we might 
expect that 
some combination of the scalar fields  $G^{ij}_n$ to couple to the SM. Here in 
fact one KK tower of scalars does couple to $\sim T_\mu^{\mu}/\mpl$. However, since 
$T_\mu^{\mu}=0$ for massless particles (except for anomalies) this coupling is 
rather small for most SM fields except for top quarks and massive 
gauge bosons. Thus, under 
most circumstances, these scalar contributions to various processes 
are rather small. ($iv$) Though each of the 
$G^{\mu\nu}_n$ are rather weakly coupled there are a {\it lot} of them and their 
density of states is closely packed. This is very important when performing sums 
over the graviton KK tower as we will see below. 

How would ADD EDs appear at colliders? Essentially, there are two important signatures 
for ADD EDs and there has been an enormous amount of work on 
the phenomenology of the ADD model in the literature [For a review see {\cite {JM}}]. 
The first signature is the emission of graviton KK tower states 
during the collision of two SM particles. Consider, \eg, either the collision of 
$q \bar q$ to make a gluon or $e^+e^-$ to make a photon and during either process 
have the SM fields emit a tower KK graviton states.  Note that since 
each of the graviton KKs  
is very weakly coupled this cross section is quite small for any given KK state. 
Also, once emitted, the 
graviton interacts so weakly it will not scatter or decay in the detector and will thus 
appear only as missing energy. Now apart from their individual 
masses all graviton KK states 
will yield the same cross section as far as this final state is concerned, \ie, 
jet or photon plus missing energy; thus we should sum up all the contributions of the KK 
states that are kinematically accessible. For example, at an $e^+e^-$ collider 
this means we 
sum over the contributions of all KK states with masses less than $\sqrt s$. This 
is a {\it lot} of states and, since these states are closely packed, we can replace 
the sum with an integral over the appropriate 
density of states. While {\it no one} KK graviton state yields a 
large cross section the resulting sum over so many KKs 
does yield a potentially large rate for either 
of the processes above. These rates only depends on the values of $n$ and $M_*$. 
As reviewed by Landsberg in these proceedings{\cite {greg}}, 
both the Tevatron and LEPII 
have looked for such signatures with no luck and have placed bounds on the ADD model 
parameters{\cite {JM}}. Clearly it is up to 
searches at future colliders to find these signals if they exist. 

\begin{figure}[htbp]
\centerline{
\includegraphics[width=10.0cm,angle=0]{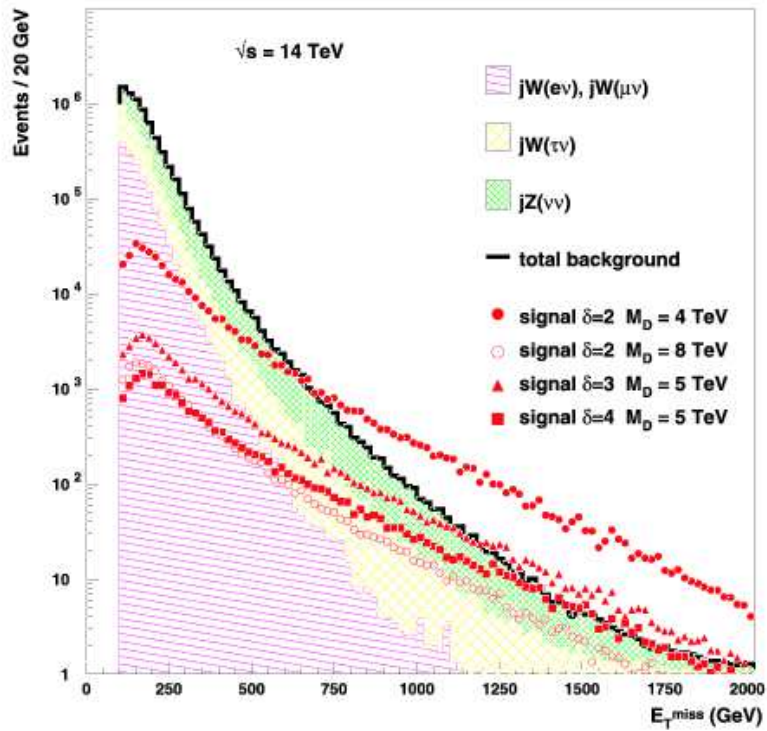}}
\vspace*{0.0cm}
\caption{Missing transverse energy spectrum for the monojet plus missing $E_T$ 
signature at the LHC assuming an integrated luminosity of 100 fb$^{-1}$ from 
Ref{\cite {vh}}. Both the 
SM backgrounds and the signal excesses from graviton emission in the ADD model 
are shown. Here $M_D=M_*$ and $\delta=n$.}  
\label{fig2}
\end{figure}

Fig.~\ref {fig2} from Vacavant and Hinchliffe{\cite {vh}} shows the missing $E_T$ 
spectrum at the LHC assuming an integrated luminosity of 100 fb$^{-1}$ for 
the process $pp\to$ jet plus missing energy in the SM and the excess induced 
by ADD graviton emission assuming different values of 
$n=\delta$ and $M_*=M_D$. Once the rather large SM 
backgrounds are well understood this excess will be clearly visible. The more 
difficult question to address is whether such an excess if observed at the LHC would 
naturally be interpreted as arising from ADD EDs as other new physics can lead to 
the same apparent final state. 

\begin{figure}[htbp]
\centerline{
\includegraphics[width=10.0cm,angle=0]{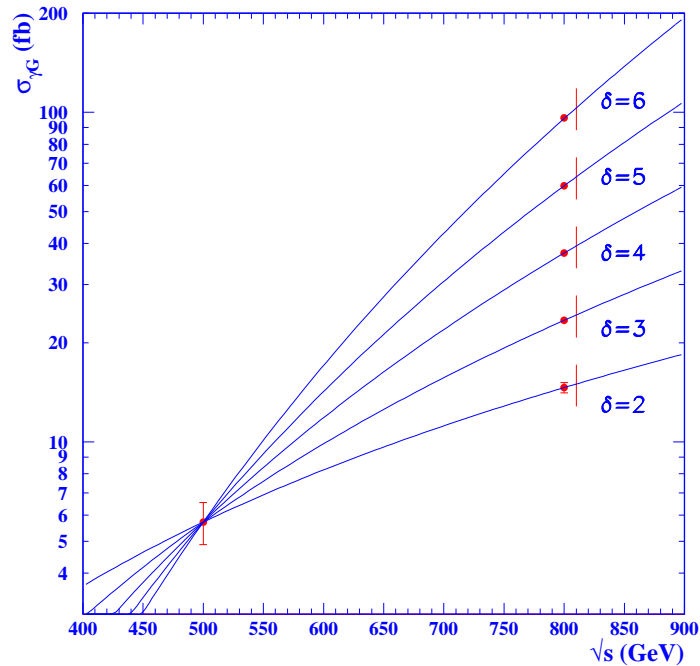}}
\vspace*{0.0cm}
\caption{Signal cross section for the $\gamma$ plus missing energy final state at the 
ILC in the ADD model 
as a function of $\sqrt s$ for various $\delta=n$ from Ref.{\cite {tdr}} normalized 
to a common value at $\sqrt s=500$ GeV. Combining measurements at two distinct values 
of $\sqrt s$ one can extract both the values of $n$ and $M_*$ for the ADD model.}
\label{fig3}
\end{figure}

At the ILC, the backgrounds for the photon plus missing energy process are far simpler 
and better understood, essentially arising from the $\nu \bar \nu+\gamma$ 
final state. These backgrounds can be measured directly by modifying the 
electron (and positron) beam 
polarization(s) since the $W^+W^-$ intermediate state gives the dominant contribution 
to this process. Measuring the excess event cross section at two different center 
of mass energies allows us to determine both $M_*$ and $n=\delta$ as shown in 
Fig.~\ref {fig3} from the TESLA TDR{\cite {tdr}}. If fitting the data taken at different 
center of mass energies results in a poor $\chi^2$ using these parameters we will 
know that the photon plus missing energy excess is due to 
some other new physics source and not to ADD EDs.

\begin{figure}[htbp]
\centerline{\includegraphics[width=9.5cm,angle=-90]{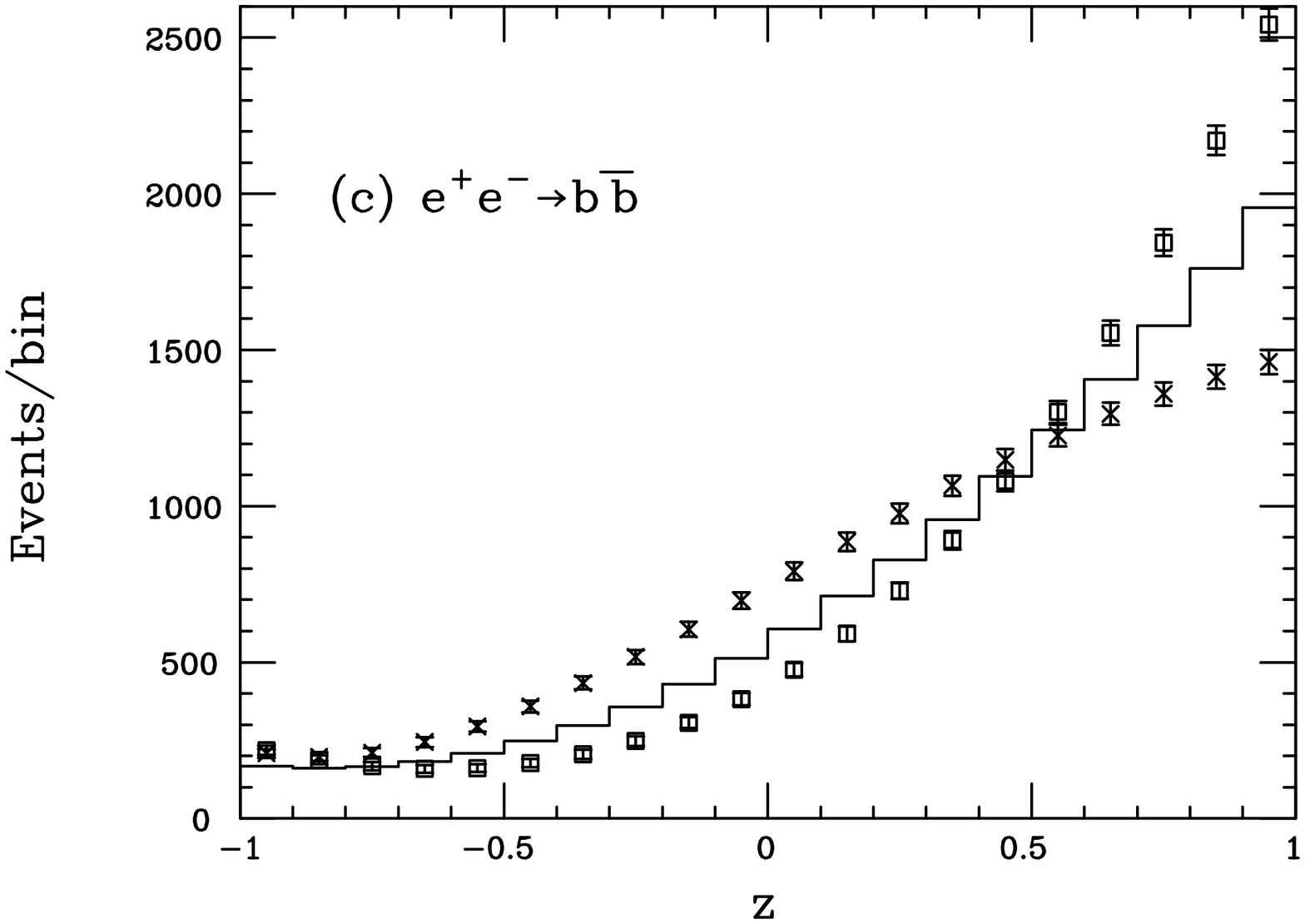}}
\vspace{-0.50cm}
\centerline{
\includegraphics[width=9.5cm,angle=-90]{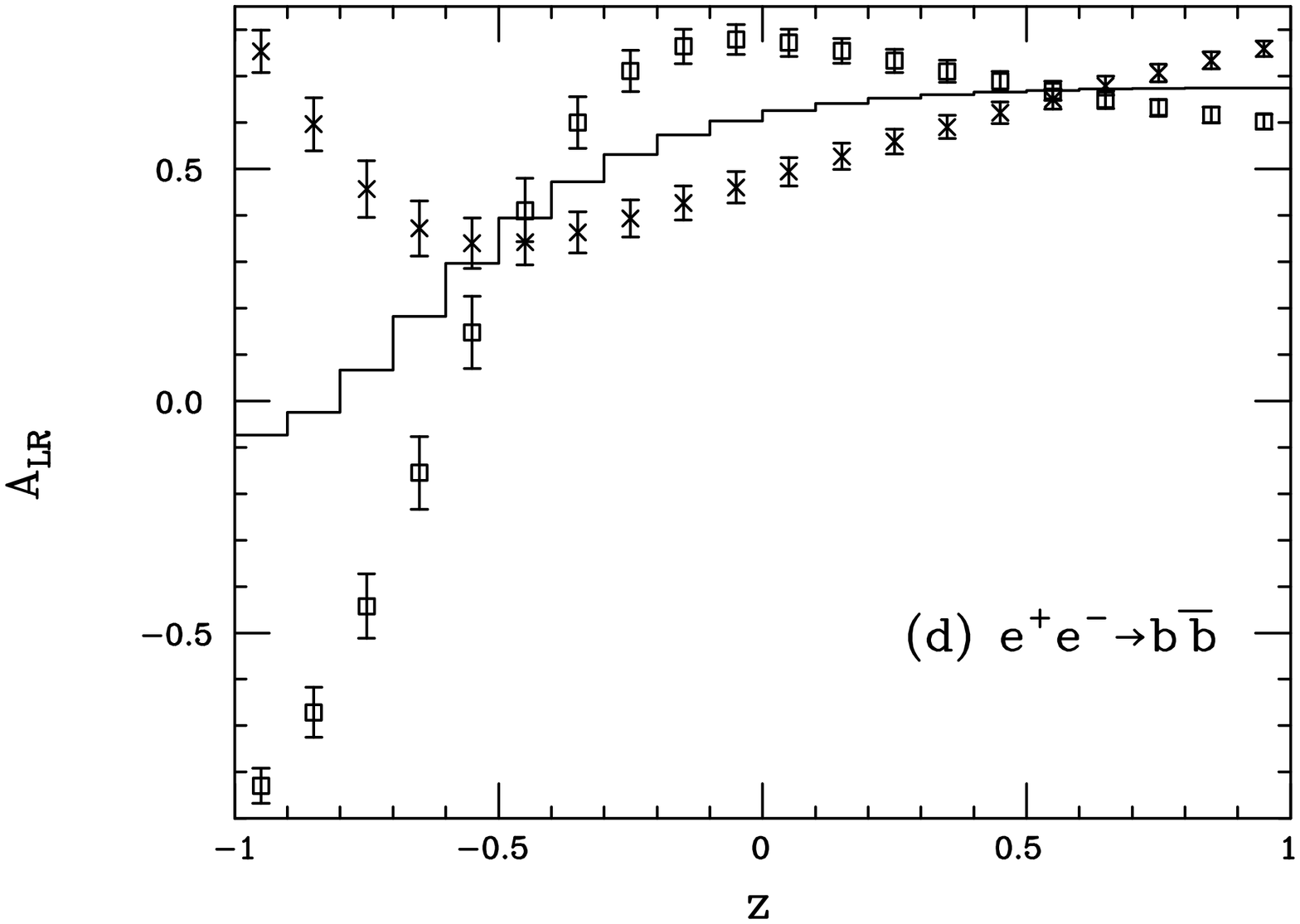}}
\vspace*{-0.50cm}
\caption{Deviations in the process $e^+e^-\to b\bar b$ at the ILC due to graviton 
KK tower exchange in the ADD model from Hewett{\cite {jlhmp}}. The top panel is 
the angular distribution while the lower panel is the left-right polarization 
asymmetry. Here $\sqrt s=500$ GeV and $\Lambda_H=1.5$ TeV. 
The histograms are the SM predictions while the `data' points are for the ADD model 
with $\lambda=\pm 1$. An integrated luminosity of 75 fb$^{-1}$ has been assumed.}
\label{fig4}
\end{figure}

These is another way to see, at least indirectly, the effect of graviton KKs in the 
ADD model: gravitons can be exchanged between colliding SM particles. This means that 
processes such as $q\bar q \to gg$ or $e^+e^- \to \mu^+\mu^-$ can proceed through 
graviton KK tower exchange as well as through the usual SM fields. As before, the 
amplitude for one KK intermediate state is quite tiny but we must again sum over all 
their exchanges (of which there are very many) thus obtaining a potentially 
large result. Unlike 
the case of graviton emission where the KK sum was cut off by the kinematics here there 
is no obvious cutoff and, in principle, the KK sum should include 
all the tower states. 
One problem with this is that this KK sum is divergent once $n>1$ as is the 
case here. (In 
fact the sum is log divergent for $n=2$ and power law divergent for larger $n$.) The 
conventional approach to this problem 
is to remember that once we pass the mass scale $\sim M_*$ the 
gravitons in the ADD model become strongly coupled and we can no longer rely on 
perturbation theory so perhaps we should cut off the sum near $M_*$. There are 
several ways to implement 
this in detail described in the literature{\cite {hlzgrw,jlhmp}}. In all 
cases the effect of graviton exchange is to produce a set of dimension-8 operators 
containing SM fields, \eg, in the notation of Hewett{\cite {jlhmp}}
\begin{equation}
{\cal L}={{4 \lambda }\over {\Lambda_H^4}} T_{\mu\nu}^i T^{\mu\nu}_f\,,
\end{equation}
where $\Lambda_H \sim M_*$ is the cutoff scale, $\lambda=\pm 1$ and $T^{\mu\nu}_{i,f}$ 
are the stress energy tensors for the SM fields in the initial and final state,  
respectively. This is just a contact interaction albeit of dimension-8 and with an 
unconventional tensor structure owing to the spin-2 nature of the gravitons being 
exchanged. 
Graviton exchange contributions to SM processes can lead to substantial 
deviations from conventional expectations; 
Fig~\ref{fig4} shows the effects of graviton KK exchange on the process 
$e^+e^-\to b\bar b$ at the ILC. Note that the differential cross section as well as 
the left-right polarization asymmetry, $A_{LR}$, are both altered from the usual SM 
predictions. 

Can the effects of graviton exchange be uniquely identified, \ie, separated from other 
new physics which induces contact interaction-like effects, such as $Z'$ 
exchange? This has been addressed by several groups of 
authors{\cite {osme}}. For example, by 
taking moments of the $e^+e^-\to f\bar f,W^+W^-$ angular distributions and employing 
polarized 
beams it is possible to uniquely identify the spin-2 nature of the graviton KK 
exchange up to $\sim 6$ TeV at a $\sqrt s=1$ TeV ILC with an integrated luminosity of 
1 ab$^{-1}$. This is about half of the discovery reach at the ILC 
for ADD EDs: $\Lambda_H \simeq 10-11\sqrt s$ for reasonable luminosities. 
If both beams could be polarized this could be improved somewhat by also 
employing transverse polarization asymmetries.

It is possible to constrain the ADD model in other ways, \eg, the emission of ADD KK 
gravitons can be constrained by astrophysical processes as 
reviewed in Ref.{\cite {JM}}. These 
essentially disfavor values of $M_*$ less than several hundred TeV for $n=2$ but yield 
significantly weaker bounds as $n$ increases. 

Before turning to a different model let us briefly discuss the dirty little secret of 
the ADD model. The purpose of this model was to eliminate the hierarchy problem, \ie, 
remove the large ratio between the weak scale and the true fundamental scale, hence 
the requirement that $M_*\sim$ a few TeV. However, if we look carefully we see that 
this large ratio has been eliminated in terms of another large ratio, \ie, $RM_* \sim 
(\mpl^2/M_*^2)^{1/n}$, which for smallish $n$ is a very large number--as large as 
the hierarchy we wanted to avoid. Thus we see that 
ADD really only trades one large ratio for another and does not really eliminate the 
hierarchy problem. The next model we will discuss does 
a much better job in that regard.

\section{Warped Extra Dimensions}

The Warped Extra Dimensions scenario was created by Randall and 
Sundrum(RS){\cite {RS}} and 
is quite different and more flexible than the ADD model. The basic 
RS model assumes the 
existence of only one ED which is compactified on the now-familiar $S^1/Z_2$ orbifold 
discussed above. In this setup there are two branes, one at $y=0$ 
(called the Planck brane) 
while the other is at $y=\pi r_c$ (called the TeV or SM brane) 
which are the two orbifold fixed 
points. What makes this model special is the metric:
\begin{equation}
ds^2=e^{-2\sigma(y)}\eta_{\mu\nu}dx^\mu dx^\nu- dy^2\,,
\end{equation}
where $\eta_{\mu\nu}=diag(1,-1,-1,-1)$ is the usual Minkowski 
metric and $\sigma(y)$ is 
some a priori unknown function. This type of geometry is called `non-factorizable' 
because 
the metric of the 4D subspace is $y-$dependent. In the simplest 
version of the RS model it is assumed, like in the ADD case, that the SM 
fields live on the so-called TeV brane while 
gravity lives everywhere. Unlike in the ADD case, however, there is a `cosmological' 
constant in the 5D bulk and both branes have distinct tensions. 
Solving the 5D Einstein's 
equations provides a unique solution for these quantities and also determines that 
$\sigma=k|y|$, where $k$ is a dimensionful parameter. 
A basic assumption of this model is 
that there are no large mass hierarchies present so that very roughly we expect 
that $k\sim M_*$, the 5D fundamental or Planck scale. In fact, once 
we solve Einstein's 
equations and plug the solutions back into the original action and 
integrate over $y$ we find that 
\begin{equation}
\mpl^2={M_*^3\over {k}}(1-e^{-2\pi kr_c})\,. 
\end{equation}
As we will see below the {\it warp factor} $e^{-\pi kr_c}$ 
will be a very small quantity 
which implies that $\mpl,M_*$ and $k$ have essentially comparable magnitudes following 
from the assumption that no hierarchies exist. If we 
calculate the Ricci curvature invariant for this 5D space 
we find it is a constant, \ie, 
$R_5=-20k^2$ and thus $k$ is a measure of the constant curvature 
of this space. A space 
with constant negative curvature is called an Anti-DeSitter space 
and so this 5D version 
is called $AdS_5$. Due to the presence of the exponential warp factor this space is 
also called a {\it warped}  space. On the other hand, a space with 
a constant metric is called `flat'; the ADD model is an example 
of flat EDs when compactified on $T^n$ as has been assumed here. (The $T^n$ spaces are 
flat since one can make a conformal transformation to a metric with constant 
coefficients.)  
Before going further we note that if the scale of curvature is too 
small, \eg, if the inverse radius of curvature becomes larger 
than the 5D Planck scale,  
then quantum gravity effects can dominate our 
discussion and our whole scenario may break 
down since we are studying the model in its `classical', \ie, non-quantum limit. 
This essentially means that we must require $|R_5| \leq M_*^2$ which 
implies a bound that $k/\mpl \leq 0.1$ or so, which is not much of a hierarchy. 

Now for the magic of the RS model. In fitting in with the RS philosophy 
it will be assumed 
that all dimensionful parameters in the action will have their mass scale set by 
$M_*\sim \mpl \sim k$ 
so that there is no fine-tuning. However, the warp factor rescales them 
as one moves about in $y$ so that, in particular, all masses 
will appear to be of order 
the TeV scale on the SM brane, \ie, to us. This means that if there is some mass 
parameter, $m$, in the action which is order \mpl, we on TeV brane will measure it to 
be reduced by the warp factor, \ie, $me^{-\pi kr_c}$. Note that if $kr_c\sim 11$ 
(a small hierarchy) this 
exponential suppression reduces a mass of order $10^{18}$ GeV to only 1 TeV. Thus the 
ratio of the weak scale to \mpl ~is explained through an exponential 
factor and no large 
ratios appear anywhere else in the model. It has been shown by Goldberger and 
Wise{\cite {gw}} that values of $kr_c \sim 11$ are indeed natural and can be provided  
by a stable configuration. Hence we have obtained a true solution 
to the hierarchy problem. 

\begin{figure}[htbp]
\centerline{\includegraphics[width=8.0cm,angle=90]{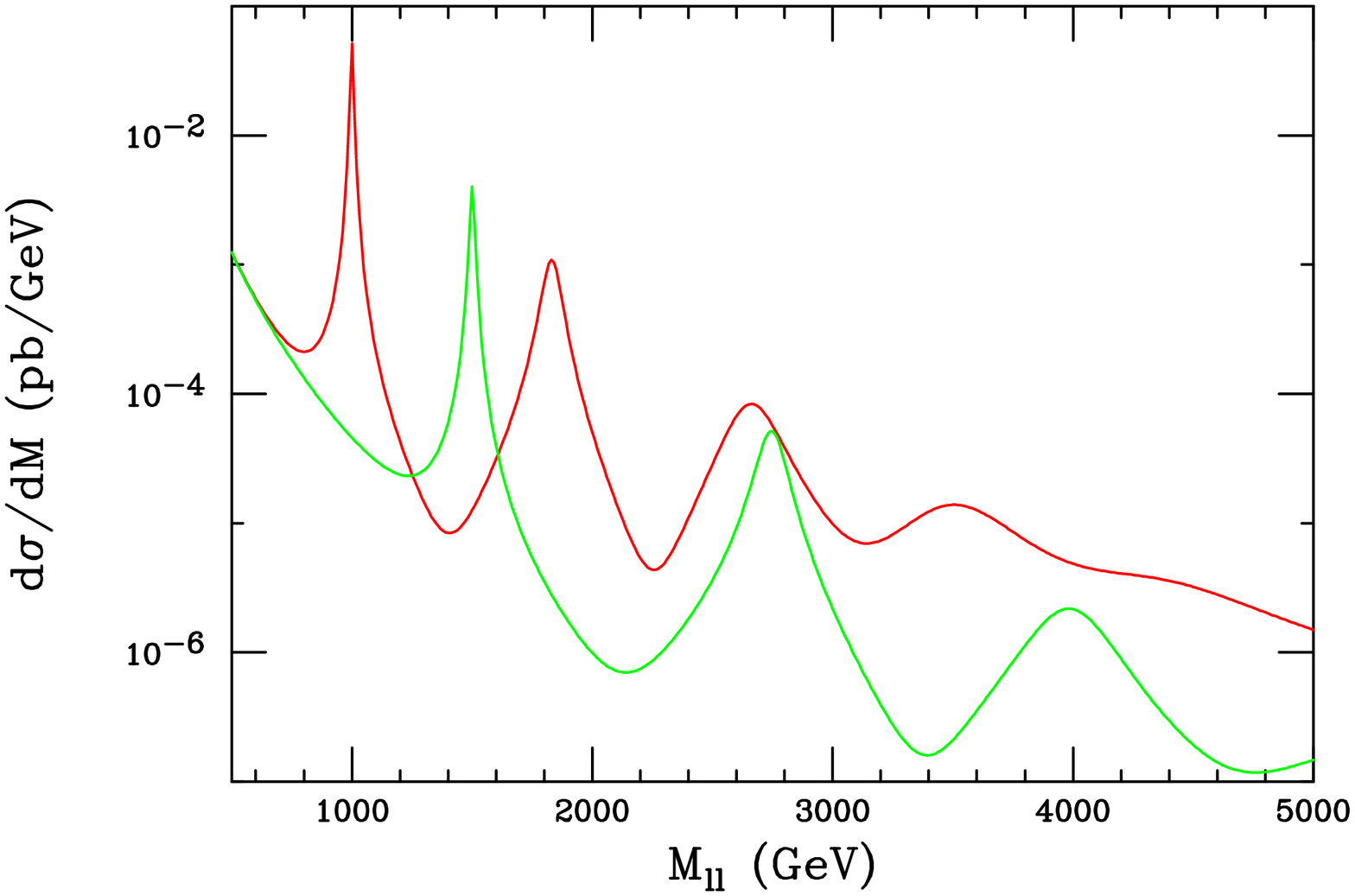}}
\vspace{0.30cm}
\centerline{
\includegraphics[width=8.0cm,angle=90]{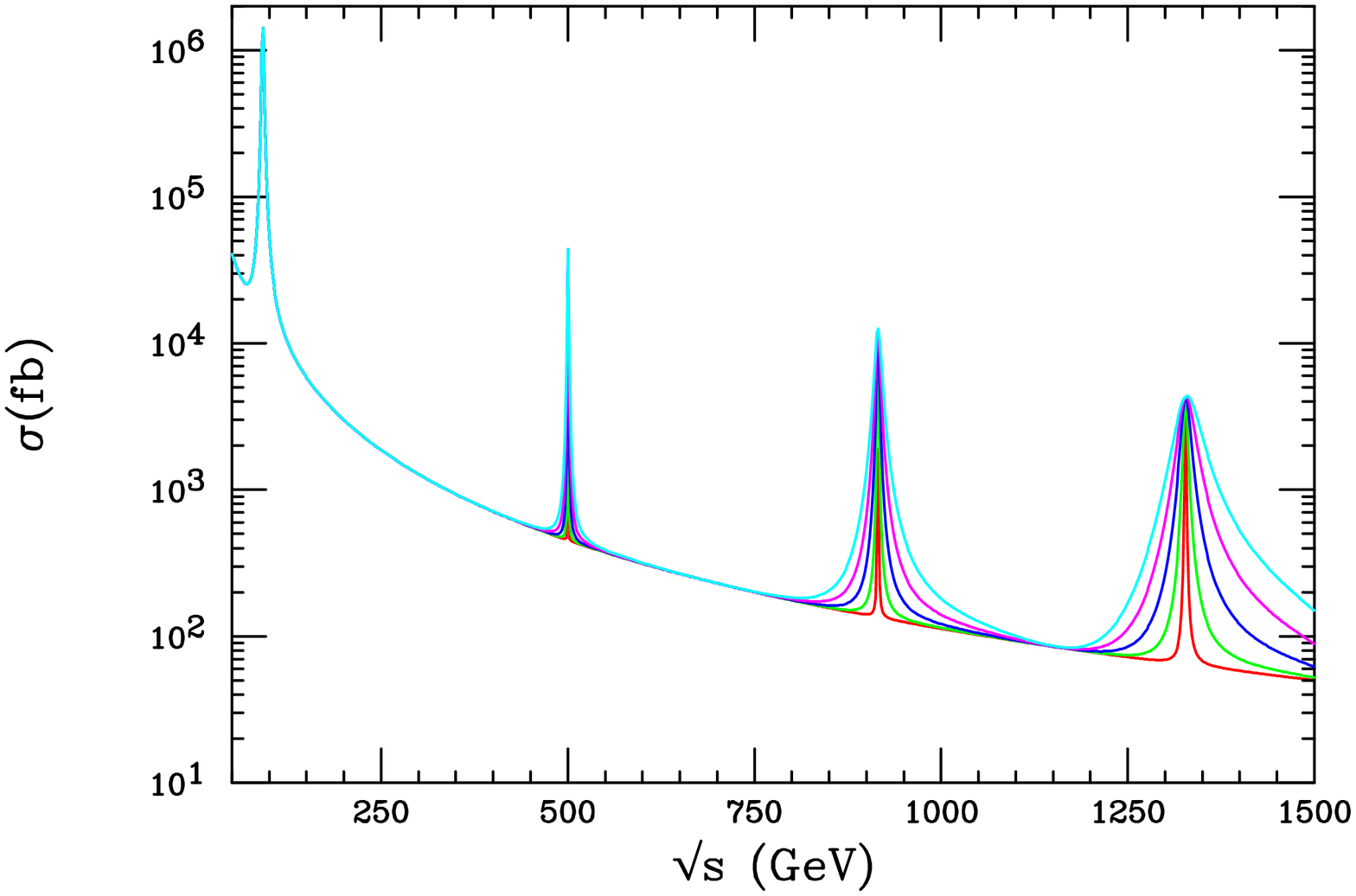}}
\vspace*{0.0cm}
\caption{Graviton resonance production in Drell-Yan at the LHC(top) and at the ILC in 
$\mu^+\mu^-$ in the RS model from Ref{\cite {us}}. The different curves correspond to 
various choices of $k/\mpl$ and $m_1$ as described in the text.The ever widening 
resonances correspond to increasing the value of $k/\mpl$.}
\label{fig5}
\end{figure}

How does this `warping' really work? Let's do a simple example 
by considering the action 
for the Higgs field on the TeV brane: 
\begin{equation}
S=\int d^4x dy ~\sqrt {-g} \big (g^{\mu\nu} \partial_\mu H^\dagger \partial_\nu H -\lambda 
(H^2-v_0^2)^2 \big ) \delta(y-\pi r_c)\,, 
\end{equation}
where $g$ is the determinant of the metric tensor, 
$\lambda$ is the usual quartic coupling 
and $v_0$ is the Higgs vev, which, keeping with the RS philosophy, is assumed to be of 
order \mpl. Now $\sqrt {-g}=e^{-4k|y|}$ and $g^{\mu\nu}=e^{2k|y|}\eta^{\mu\nu}$ 
so that 
we can trivially integrate over $y$ due to the delta-function. This yields 
\begin{equation}
S=\int d^4x \big (e^{-2\pi kr_c} \eta^{\mu\nu} 
\partial_\mu H^\dagger \partial_\nu H -\lambda e^{-4\pi kr_c} 
(H^2-v_0^2)^2 \big)\,. 
\end{equation}
Now to get a canonically normalized Higgs field 
(one with no extra constants in front of the 
kinetic term) we rescale the field by letting $H\to e^{\pi kr_c}h$ which now gives 
\begin{equation}
S=\int d^4x \big (\partial^\mu h^\dagger \partial_\mu h -\lambda (h^2-v_0^2 e^{-2\pi kr_c})^2
\big )\,, 
\end{equation}
where we have contracted the indices using $\eta^{\mu\nu}$. 
We see that the vev that we observe on the SM brane is 
{\it not} $v_0$ but $v_0e^{-\pi kr_c}$ 
which is of order the TeV scale. Thus the Higgs gets a TeV scale vev even though the 
parameters we started 
with in the action are all of order \mpl! This warping effect is a 
general result of the RS model.

\begin{figure}[htbp]
\centerline{
\includegraphics[width=9.0cm,angle=90]{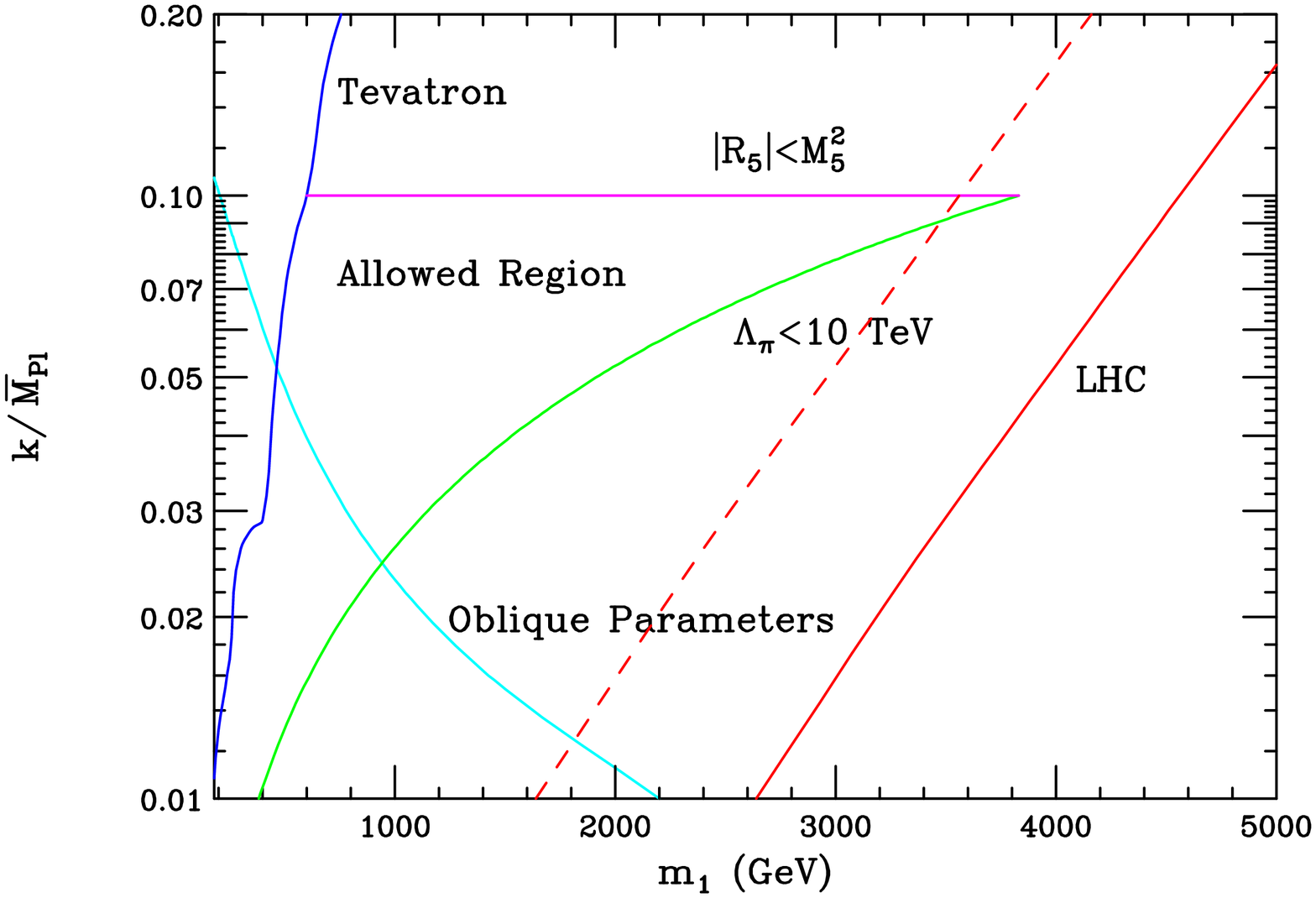}}
\vspace*{0.0cm}
\caption{Allowed region in the RS model parameter space 
implied by various theoretical and 
experimental constraints from Ref.{\cite {us}}. 
The regions to the left of the horizontal 
lines are excluded by direct searches at 
colliders. The dashed(solid) line for the LHC corresponds to an 
integrated luminosity of 10(100) fb$^{-1}$. The present anticipated parameter space is 
inside the triangular shaped region.} 
\label{fig6}
\end{figure}

What do the gravitons look like in this model? Even though gravitons 
are spin-2 it turns 
out that their masses and wave functions are identical to the case of a 
scalar field in 
the RS bulk{\cite {dhr}} which is far simpler to analyze. Let us return to the 
Klein-Gordon equation above but now in the case 
of curved space; one obtains 
\begin{equation}
(\sqrt {-g})^{-1}\partial_A \big (\sqrt {-g} g^{AB} \partial_B \Phi \big )=0\,. 
\end{equation}
`Separation of variables' via the KK decomposition then yields 

\begin{equation}
-e^{2ky} \partial_y \big (e^{-2ky}\partial_y \chi_n\big )=m_n^2\chi_n\,, 
\end{equation}
which reduces to the result above for a space of 
zero constant curvature, \ie, when $k\to 0$. 
The solutions to this equation for the $\chi_n$ 
wave functions yield linear combinations of the $J_2,Y_2$ Bessel functions 
and not sines and cosines as in the flat space case and the masses of the 
KK states are given by 
\begin{equation}
m_n=x_nke^{-\pi kr_c}\,, 
\end{equation}
where the $x_n$ are roots of $J_1(x_n)=0$. Here 
$x_n= 0,~3.8317..,~7.0155..,~10.173..,..$ 
\etc. Since $ke^{-\pi kr_c}$ is $\sim$ a few hundred GeV at most, we see that 
the KK graviton masses are of 
a similar magnitude with comparable, but unequal, spacing, \ie, the KK gravitons have 
approximately weak/TeV scale masses. This is quite different than in the ADD model. 
Returning to the 5D Einstein action we can insert 
the wavefunctions for the KK states and 
determine how they couple to SM fields on the TeV brane; one finds 
\begin{equation}
{\cal L}=-\Big ({{G^{\mu\nu}_0} \over {\mpl}} +\sum_{n>0} 
{{G^{\mu\nu}_n}\over {\Lambda_\pi}} 
\Big) T_{\mu\nu}\,, 
\end{equation}
where $\Lambda_\pi=\mpl e^{-\pi kr_c}$ is of order TeV. Here we see that the ordinary 
graviton zero mode couples as in the ADD model as it should 
but all the higher KK modes have couplings 
that are exponentially larger due to the common warp factor. 
We thus have weak scale graviton 
KKs with weak scale couplings that should be produced as spin-2 {\it resonances} at 
colliders. Due to the universality of gravity these KK graviton resonances 
should be observable in many processes. There are no table top or 
astrophysical constraints on this scenario unlike in the ADD model. 
It is interesting to note that 
this model also has only 2 free parameters which we can take to be the mass of the 
lightest KK excitation, $m_1$, and the ratio $k/\mpl$; given these parameters as  
input all other masses and 
couplings can be determined. As we will see $k/\mpl$ essentially 
controls the KK resonance width for a fixed value of the resonance mass.  
The RS model in its simplest form is thus highly predictive. 
\begin{figure}[htbp]
\centerline{\includegraphics[width=10.0cm,angle=0]{4arnps_spin_grav.epsi}}
\vspace{0.30cm}
\centerline{
\includegraphics[width=8.0cm,angle=90]{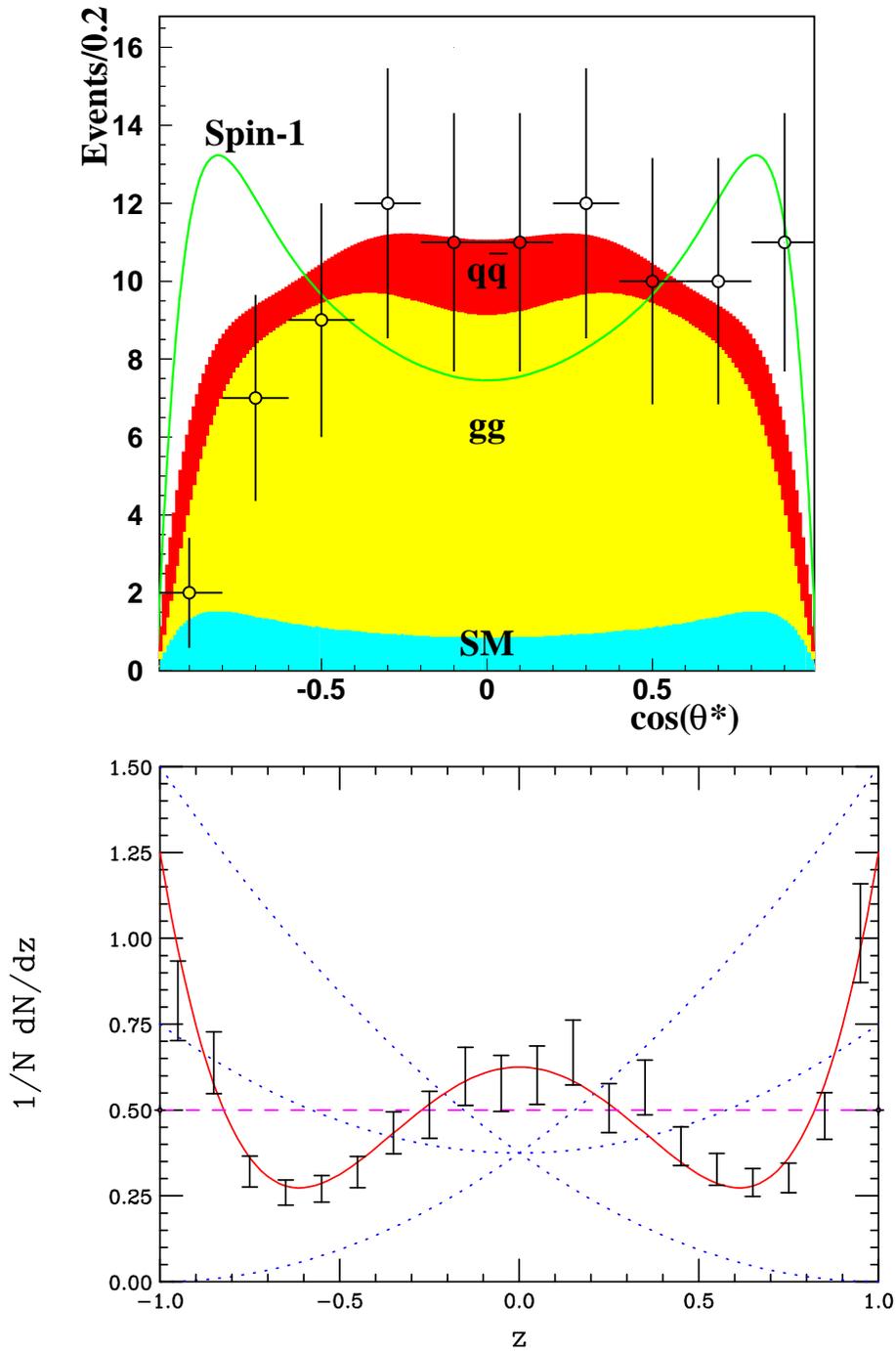}}
\vspace*{0.0cm}
\caption{Results from Refs.{\cite {brits,us}} showing that the spin of the KK graviton 
in the RS model can be determined at either the LHC(top) or ILC(bottom) 
from the angular distribution of final state dilepton pairs. Fitting the dilepton 
data to different spin hypotheses is relatively straightforward.}
\label{fig7}
\end{figure}

At this point one may wonder why in the RS model the 
zero-mode coupling is so weak while the 
couplings of all the other KK tower states are so much stronger. 
The strength of the graviton 
KK coupling to any SM state on the TeV brane is proportional to the value of its 5D 
wavefunction at $y=\pi r_c$.  In the flat space cases discussed above the 
typical wavefunctions for 
KK gravitons were $\sim \cos ny/R$ and so took on essentially the same 
value at the location of the SM fields for all $n$. Here 
the relevant combinations of the 
$J_2,Y_2$ Bessel functions behave quite differently when $x_n=0$, 
\ie, for the zero mode, 
versus the case when $x_n$ takes on a non-zero value as it does for the 
KK excitations. For the 
zero mode the 5D wavefunction is highly peaked near the Planck 
brane and so its value is very small near 
the TeV brane where we are; the opposite is true for the other KK states. 
Thus it is the strong 
peaking of the graviton wavefunctons that determine the strength of the gravitational 
interactions of the KK states with us.

What will these graviton KK states look like at a collider? 
Fig.~\ref{fig5} shows the production 
of graviton resonances at the LHC in the Drell-Yan channel and in $\mu$-pairs 
at the ILC for 
different values of $m_1$ and $k/\mpl${\cite {dhr,us}}. 
Note that the width of the resonance grows as $\sim (k/\mpl)^2$ 
so that the resonance appears rather like a spike when this ratio is small. 
Also note that due to the 
nonrenormalizable coupling of the graviton KK states the resonance 
width also grows as $\Gamma \sim m_n^3$ as we 
go up the KK tower. Hence heavier 
states are rather wide; for any reasonable fixed value of 
$k/\mpl$ at some point as one goes up the KK tower one reaches states which are 
quite wide 
with $\Gamma \simeq m_n$ signaling the existence of the 
strongly interaction sector of the theory.  
Examining the $k/\mpl-m_1$ parameter space and making some simple assumptions 
one sees that the LHC has a very good chance of 
covering all of it once 100 fb$^{-1}$ of integrated luminosity have 
been accumulated. Part of 
the present constraints on RS follow from the requirement that 
$k/\mpl \leq 0.1$, as discussed 
above, and how large we are willing to let $\Lambda_\pi$ be before we 
start worrying about fine tuning again. Given these considerations we see that the 
LHC has excellent RS parameter space 
coverage as seen in  Fig.~\ref{fig6}. 

\begin{figure}[htbp]
\centerline{
\includegraphics[width=8.5cm,angle=90]{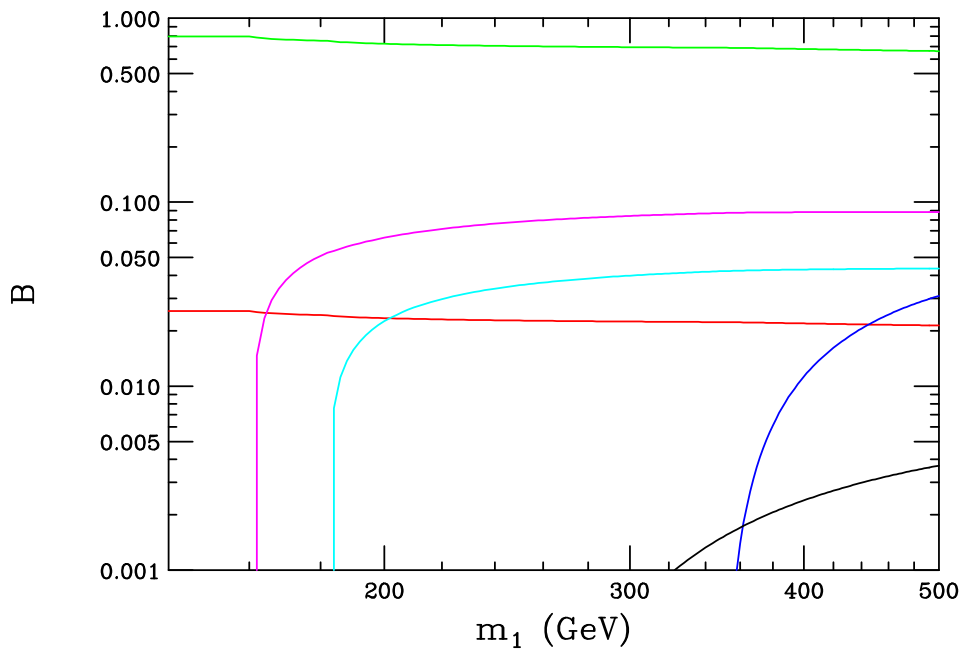}}
\vspace*{0.0cm}
\caption{Branching fractions for the RS graviton KK state as a function 
of its mass from 
Ref.{\cite {us}}.  From top to bottom on the right hand side of the figure the curves 
correspond  to the following final states: $jj$, $W^+W^-$, $ZZ$, $t\bar t$, 
$\ell^+\ell^-$, and $hh$, respectively.}
\label{fig8}
\end{figure}

Once we discover a new 
resonance at the LHC or ILC we'd like to know whether or not it is a 
graviton KK state. The first thing to do is to 
determine the spin of the state; Fig.~\ref{fig7}  
shows that differentiating spin-2 from other possibilities is relatively 
straightforward{\cite {brits,us}} at 
either machine. To truly identify these spin-2 resonances as 
gravitons, however, we need to 
demonstrate that they couple universally as expected from General Relativity. 
The only way to do this is 
to measure the various branching fractions and this is most easily done 
at the ILC. Fig.~\ref{fig8} shows 
the expected branching fractions for a graviton KK as a 
function of its mass assuming only 
decays to SM particles with 
a Higgs mass of 120 GeV. One unique test {\cite {us}} is based 
on the fact that $\Gamma(G_n\to \gamma \gamma)=2\Gamma(G_n\to \ell^+\ell^-)$ 
both of which can be easily measured at either collider. 

Before concluding this section we should note that the 
simple RS model scenario is barely the 
tip of the iceberg and has been extended in many 
ways to help with various model building 
efforts. A few possibilities that have been considered (with limited references!) are
\begin{itemize}
\item  Extend to 3 or more branes{\cite {new1}}
\item  Extend to 6 or more dimensions{\cite {new2}} 
\item  Put the SM gauge fields and fermions in the bulk{\cite {big}} with or without 
       localized brane term interactions{\cite {brane}}. This is very active are 
       of current research. 
\end{itemize}

\section{Universal Extra Dimensions}

\begin{figure}
\includegraphics[width=.48\textwidth]{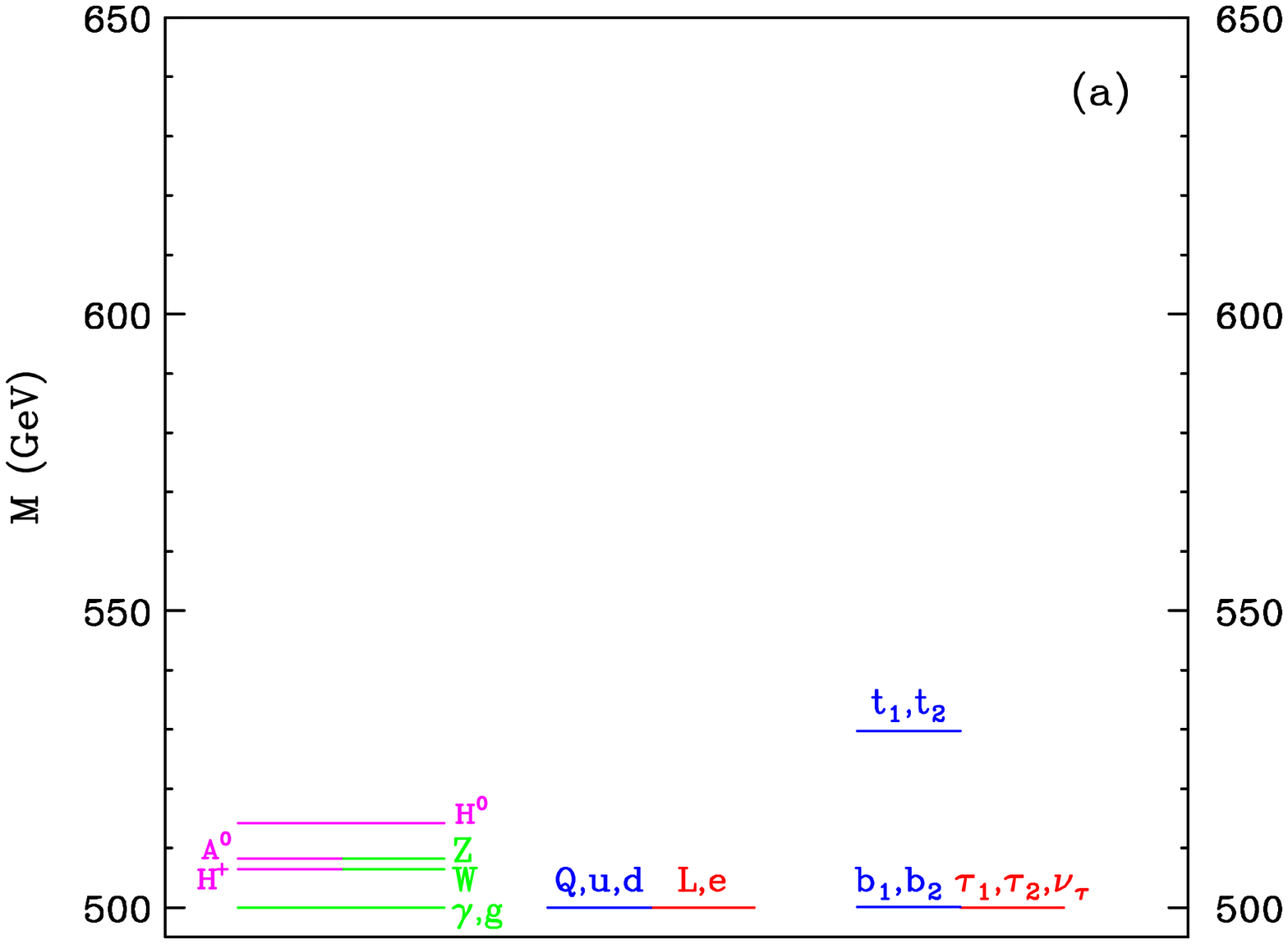}
\includegraphics[width=.48\textwidth]{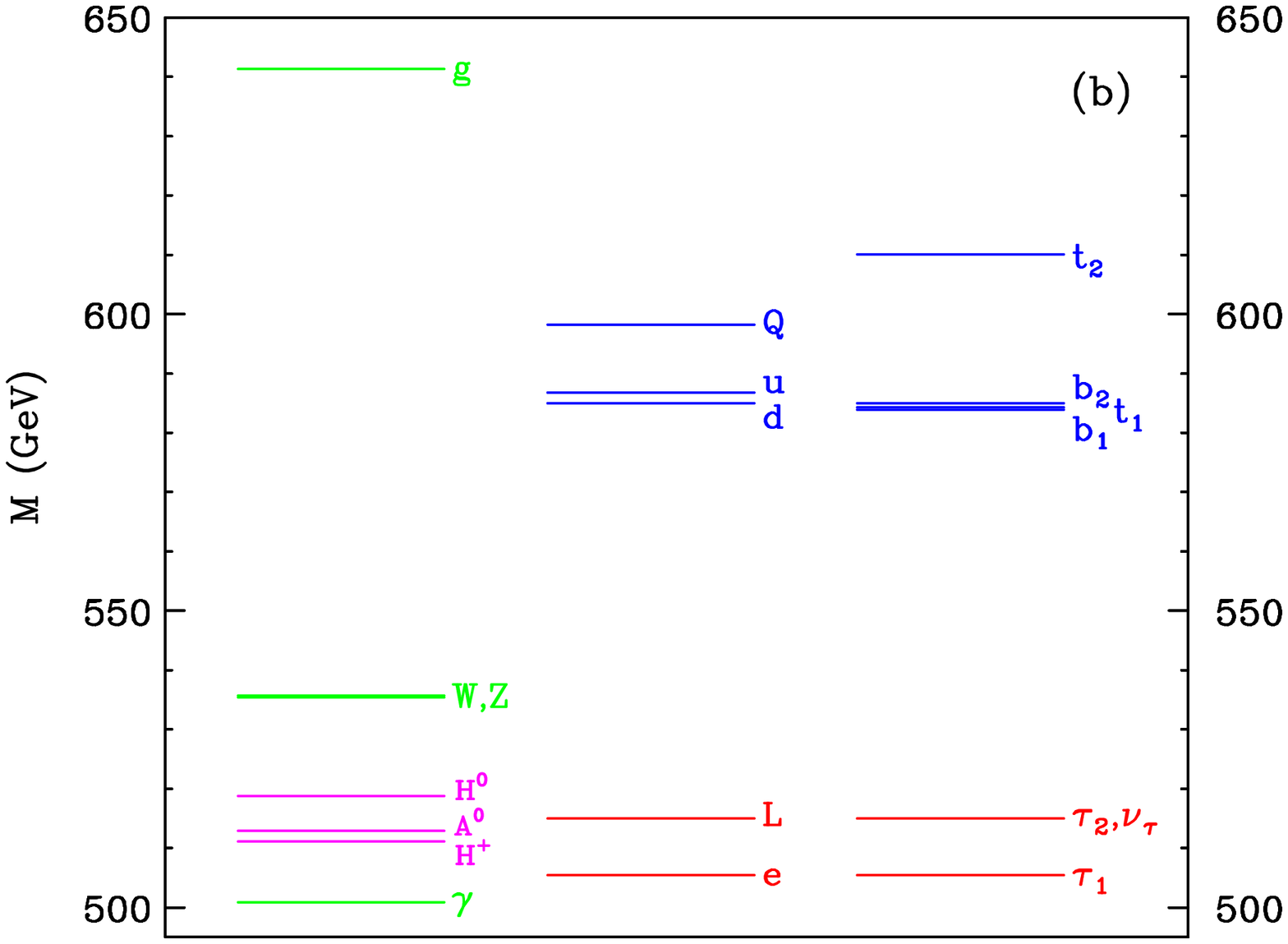}
\caption{\label{fig:spectrum} The spectrum of the first KK level
at (a) tree level and (b) one-loop, for
$R^{-1}=500$ GeV, $\Lambda R = 20$, $m_h=120$ GeV,
$\overline{m}_H^2=0$, and
assuming vanishing boundary terms at the cut-off scale $\Lambda$ from Cheng, 
Matchev and Schmaltz{\cite {ued}}.}
\label{fig9}
\end{figure}

Let us return to the case where we again have one flat 
TeV-scale ED as we might have in 
ADD with different compactification radii as discussed above. 
Then we can imagine putting 
all of the SM fields into this part of the bulk so that they have KK 
excitations; this is the Universal 
Extra Dimensions(UED) scenario of Appelquist, Cheng and 
Dobrescu{\cite {ued}}. For simplicity 
we can think of this one ED as just $S^1/Z_2$ as we are familiar with from 
above. The KK masses 
of the various fields at tree level are then given by
\begin{equation}
m_{KK}^2=m_{SM}^2+n^2/R^2\,, 
\end{equation}
where $m_{SM}$ are the various SM particle masses and $R$ is 
the usual compactification radius. 
Note that even if $R^{-1}$ is only 300-500 GeV there will be a lot 
of degeneracy among the KK 
levels. The details of the KK mass spectrum are clearly 
very important for understanding the 
phenomenology of this model, \eg, the width for $e_1\to e_0\gamma_1$ is 
very sensitive to the 
details of the various masses. Thus the radiative corrections to the 
naive mass formula above 
need to be considered; these were calculated by Cheng, Matchev 
and Schmaltz{\cite {ued}} and 
they move the spectrum around by a substantial amount.  
These radiative corrections arise from two sources:
($i$) bulk terms which are finite and ($ii$) terms induced 
at the orbifold fixed points which 
depend logarithmically on a cutoff $\Lambda$. The size of such terms 
are set in the model by the assumption 
that this non-normalizable theory is replaced by some other new physics 
above the scale 
$\Lambda \simeq 20R^{-1}$ so that the magnitude of the divergent 
terms are chosen to vanish at 
the cutoff. Fig.~\ref{fig9} shows how these radiative corrections shift the tree level 
spectrum in this case. 

In a theory with only periodic BCs, such as one compactified on $S^1$, 
the value(s) of the 
momenta in the extra dimensions are conserved in any process. In the case of UED, the 
effect of radiative corrections inducing boundary terms at 
the fixed points is to violate 
this conservation law. Instead one finds a discrete conservation law, called 
KK-parity{\cite {ued}}, which takes on a value $(-1)^n$ with $n$ being the 
KK level. The 
effect of this conservation law is quite restrictive when considering the 
properties of the first KK states but is useful in helping the model avoid known 
collider and electroweak constraints. Clearly the KKs can only be pair produced 
and the lightest KK particle, the LKP, 
is stable and can be a Dark Matter candidate{\cite {ued}} for $R^{-1}$ 
in the 500-1000 GeV 
range; this sounds a lot like supersymmetry with $R$-parity conservation. 
In fact the KK 
spectrum after radiative corrections looks a lot like a typical SUSY 
spectrum. Is it possible 
that the two models could be confused{\cite {ued}}? How can they be distinguished at a 
collider? Clearly a major difference between SUSY and UED is that 
the new heavy particles have 
different spins in the two cases. Determining the spin of a new particle which is pair 
produced at the LHC is very difficult but has been 
discussed{\cite {atlnote}}; this is a much 
simpler task at the ILC as new charged spin-0 (SUSY) and spin-$1/2$ (UED) states have 
completely different threshold behaviors as discussed in Ref~{\cite {marco}}. Another 
possibility, provided $R^{-1}$ is not too large, is the single 
production of the $n=2$ states 
in the UED framework. No particles analogous to these exist in the SUSY case so if 
observed would uniquely point to the UED scenario.

\section{Summary and Conclusion}

The subject of EDs has become a huge research area over the last six years and we have 
hardly scratched the surface in the present discussion. As one can see there are at 
present an immense number of ideas and models floating around connected to EDs and 
we certainly can expect there to be many more in the future. EDs 
can lead to a wide range of new phenomena (Dark Matter, collider signatures, \etc) 
that will be sought over the coming decade. Of course, only experiment can tell us if 
EDs have anything to do with reality and, if they do exist, what their nature may be. 
The discovery of EDs will certainly radically alter our view of the universe on the 
very small and very large scales.

\begin{acknowledgments}
The author would like to thank Hooman Davoudiasl, JoAnne Hewett, Ben Lillie, Frank 
Petriello and Jim Wells for collaboration on research on EDs over the last 6 years. 
Work supported by Department of Energy contract DE-AC02-76SF00515.
\end{acknowledgments}


%%%%%%%%%%%%%%%%%%%%%%%%%%%%%%%%%%%%%%%%%%%%%%%%%%%%%%%
\def\MPL #1 #2 #3 {Mod. Phys. Lett. {\bf#1},\ #2 (#3)}
\def\NPB #1 #2 #3 {Nucl. Phys. {\bf#1},\ #2 (#3)}
\def\PLB #1 #2 #3 {Phys. Lett. {\bf#1},\ #2 (#3)}
\def\PR #1 #2 #3 {Phys. Rep. {\bf#1},\ #2 (#3)}
\def\PRD #1 #2 #3 {Phys. Rev. {\bf#1},\ #2 (#3)}
\def\PRL #1 #2 #3 {Phys. Rev. Lett. {\bf#1},\ #2 (#3)}
\def\RMP #1 #2 #3 {Rev. Mod. Phys. {\bf#1},\ #2 (#3)}
\def\NIM #1 #2 #3 {Nuc. Inst. Meth. {\bf#1},\ #2 (#3)}
\def\ZPC #1 #2 #3 {Z. Phys. {\bf#1},\ #2 (#3)}
\def\EJPC #1 #2 #3 {E. Phys. J. {\bf#1},\ #2 (#3)}
\def\IJMP #1 #2 #3 {Int. J. Mod. Phys. {\bf#1},\ #2 (#3)}
\def\JHEP #1 #2 #3 {J. High En. Phys. {\bf#1},\ #2 (#3)}

\end{document}